\documentclass[sigconf]{aamas} 

\usepackage{balance} 

\usepackage{tikz}
\usepackage[T1]{fontenc}
\usepackage{helvet}
\usepackage{adjustbox}

\usepackage{algorithm}
\usepackage{algpseudocode}
\usepackage{amsmath}

\usepackage{booktabs}
\usepackage{multirow}
\renewcommand{\algorithmiccomment}[1]{\bgroup\hfill//~#1\egroup}

\setcopyright{ifaamas}
\acmConference[AAMAS '26]{Proc.\@ of the 25th International Conference
on Autonomous Agents and Multiagent Systems (AAMAS 2026)}{May 25 -- 29, 2026}
{Paphos, Cyprus}{C.~Amato, L.~Dennis, V.~Mascardi, J.~Thangarajah (eds.)}
\copyrightyear{2026}
\acmYear{2026}
\acmDOI{}
\acmPrice{}
\acmISBN{}

\acmSubmissionID{1698}

\title[AAMAS-2026 Formatting Instructions]{Multi-Agent Cooperative Transportation: Optimal and Efficient Task Allocation and Path Finding}

\author{Ning Zhou}
\affiliation{
  \institution{University of Bristol}
  \city{Bristol}
  \country{United Kingdom}}
\email{ning.zhou@bristol.ac.uk}

\author{Nikolai W.F. Bode}
\affiliation{
  \institution{University of Bristol}
  \city{Bristol}
  \country{United Kingdom}}
\email{nikolai.bode@bristol.ac.uk}

\author{Edmund R. Hunt}
\affiliation{
  \institution{University of Bristol}
  \city{Bristol}
  \country{United Kingdom}}
\email{edmund.hunt@bristol.ac.uk}

\begin{abstract}
Multi-robot systems are integral to modern logistics, but their capabilities are often limited to tasks executable by individual agents. This paper addresses a critical gap in existing frameworks like Multi-Agent Path Finding (MAPF) and Task Allocation and Path Finding (TAPF), which lack true cooperation for transporting large items that require multiple agents. To this end, we formalise the Cooperative Transportation Task Allocation and Path Finding (CT-TAPF) problem, which integrates team formation, task assignment, and collision-free pathfinding. We present an optimal solver, Cooperative Transportation Task Conflict-Based Search (CT-TCBS), which features a novel \textit{Incremental Expansion} strategy to tackle the combinatorial explosion inherent in team formation. Recognising the computational cost of optimality, we also develop a family of sub-optimal solvers that employ a global, task-centric perspective, selecting the next task to assign based on a global difficulty metric (Best Task or Worst Task). Our comprehensive empirical evaluation demonstrates three key findings: (1) the incremental expansion strategy significantly outperforms the naive combinatorial approach by successfully pruning the dominant task-allocation search space; (2) we identify a \textit{task-conflict expansion dilemma}, where sophisticated conflict resolvers effective for large-agent pathfinding subproblems can be detrimental in the integrated CT-TAPF setting; and (3) our proposed sub-optimal solvers establish a new, more efficient frontier on the solution quality-runtime spectrum compared to  \texttt{nn-}, agent-centric baselines. This work provides a foundational framework and a set of effective algorithms for a new, practical class of cooperative multi-agent problems.
\end{abstract}


\newcommand{\BibTeX}{\rm B\kern-.05em{\sc i\kern-.025em b}\kern-.08em\TeX}

\begin{document}


\pagestyle{fancy}
\fancyhead{}


\maketitle 


\section{Introduction}
\label{sec:intro}

Multi-robot systems are increasingly central to logistics automation, with prominent examples in large-scale warehouses~\cite{Amoo2024}. However, a critical limitation in current implementations is that agents typically operate in a shared environment but in isolation, handling uniform pods sized for a single agent. This lack of cooperation restricts operational flexibility, making it difficult to transport larger or irregularly shaped goods. While Multi-Agent Path Finding (MAPF)~\cite{Stern2019b} addresses collision-free navigation and Combined Task Allocation and Path Finding (TAPF) ~\cite{Hoenig2018a,Henkel2019a} extends this to include task selection, both paradigms predominantly focus on tasks executable by individual agents. The potential synergy from agents cooperating to perform complex transportation tasks remains a significant, underexplored area. 

\begin{figure}[h]
  \centering
  \includegraphics[width=0.5\linewidth]{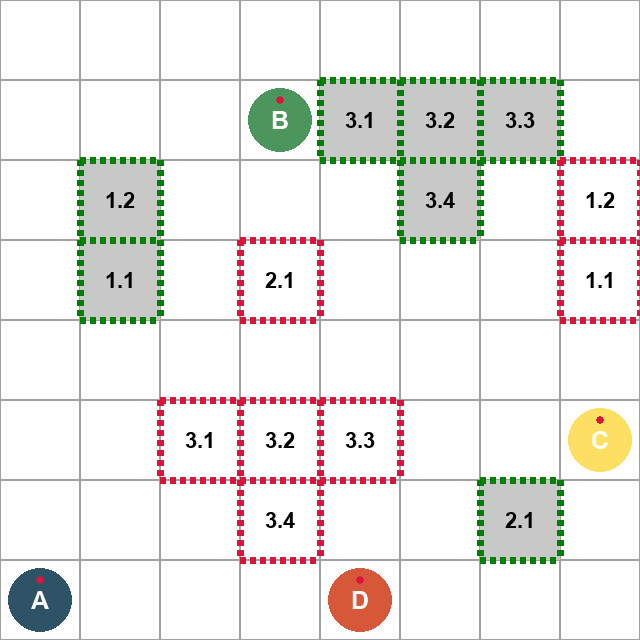}
    \caption{A CT-TAPF problem instance with four agents and three cooperative transportation tasks. Agents must form teams to move from the green dashed start locations to the red dashed goal locations. The tasks require one, two, and four agents, respectively.}
  \label{fig:ct-tapf-example}
  \Description{An 8 by 8 grid is shown, with the origin (0,0) at the bottom-left corner. There are four circular agents on the grid: Agent A is dark blue at position (0, 0). Agent B is green at (3, 6). Agent C is yellow at (7, 2). Agent D is reddish-orange at (4, 0). The grid also contains locations for three tasks. Task start locations have a green dashed border and a grey background, while goal locations have a red dashed border. Task 1, a two-agent task, has start positions labelled 1.1 at (1, 4) and 1.2 at (1, 5), with goal positions 1.1 at (7, 4) and 1.2 at (7, 5). Task 2, a single-agent task, has its start position labelled 2.1 at (6, 1) and its goal at (3, 4). Task 3, a four-agent task, has start positions labelled 3.1 to 3.4 at (4, 6), (5, 6), (6, 6), and (5, 5) respectively, with its goal positions at (2, 2), (3, 2), (4, 2), and (3, 1).}
\end{figure}

To address this gap, we formalise the Cooperative Transportation Task Allocation and Path Finding (CT-TAPF) problem, where teams of agents must be assigned to cooperative transportation tasks and execute them via collision-free paths as shown in Figure.~\ref{fig:ct-tapf-example}. As a generalisation of TAPF, the CT-TAPF problem is NP-hard. In this paper, we introduce an optimal solver, Cooperative Task Conflict-Based Search (CT-TCBS), to solve this problem. Recognising the need for scalability~\cite{Jiang2024b}, we also develop several suboptimal variants of CT-TCBS that are designed to find high-quality solutions more efficiently.

We provide a comprehensive empirical evaluation of our proposed algorithms. Our experiments measure success rates under computational constraints for optimal CT-TCBS, analyse the inherent trade-off between the search spaces for task allocation and conflict resolution, and investigate the balance between solution quality and runtime across optimal and suboptimal approaches.

We begin in Section~\ref{sec:related_work} with an overview of related work and an introduction to the CT-TAPF problem. Following this, we present the formal problem formulation in Section~\ref{sec:problemdef}. We then describe the architecture of our optimal and suboptimal solvers in Section~\ref{sec:CT-TCBS}. Finally, we evaluate the performance of our approach through simulations and present the results in Section~\ref{sec:experiment}.


\section{Related Work}
\label{sec:related_work}

The foundational problem in our domain is Multi-Agent Path Finding (MAPF), which seeks collision-free paths for agents from their unique start to goal locations. The objective is typically to minimise the sum of costs (SoC) or makespan~\cite{Stern2019b}. Both optimal and suboptimal algorithms have been proposed to solve this NP-hard problem~\cite{Yu2013f}. Optimal methods span several categories, including reduction-based algorithms ~\cite{Yu2013h,Surynek2016}, 
A*-based algorithms on coupled search space(M*~\cite{GWagn11a2011}, EPEA*~\cite{Goldenberg2014},  ODA*~\cite{Standley2010a}), and multi-level search-based algorithms (ICTS~\cite{Sharon2013}, CBS~\cite{AFeln15a}). Among optimal solvers, Conflict-Based Search (CBS) is the dominant paradigm with many modifications (e.g. ICBS~\cite{Boyarski2015a}, IDCBS~\cite{Boyarski2020a}, Adding Heuristics~\cite{Felner2018}, Symmetry Breaking~\cite{Li2020k,Li2019h}). For large-scale problems where optimality is intractable, scalable sub-optimal algorithms, such as PP~\cite{Silver2005b}, LNS2~\cite{Li2022k}, LACAM~\cite{Okumura2023}, provide practical solutions.

Real-world applications require agents to also select which task to execute, leading to the Task Assignment and Path Finding (TAPF) problem~\cite{Hoenig2018a}, a more general formulation closely related to Multi-Agent Pickup and Delivery (MAPD)~\cite{Ma2017b}. In TAPF, the tight coupling of task assignment and path planning is a significant challenge, as a locally optimal assignment can create intractable congestion. Consequently, many successful TAPF solvers extend the CBS framework to handle this additional task assignment layer, including CBS-TA~\cite{Hoenig2018a}, ITA-CBS~\cite{Tang2023b} and TCBS~\cite{Henkel2019a}.

Despite their success, MAPF and TAPF frameworks primarily model \textit{coordination} (avoiding negative interference) rather than \textit{active cooperation} (agents working together to achieve synergy).  These problem formulations are fundamentally individualistic; for instance, a warehouse system using TAPF can assign robots to retrieve individual pallets but lacks the mechanism to assign two agents to transport a single oversized pallet.

The need for more substantive, cooperative models has been recognised. Prior works have introduced abstract forms of cooperation to MAPF. For instance, Cooperative MAPF (Co-MAPF) introduces cooperation as a sequential dependency, modelling tasks that require a hand-over between agents at a specific location~\cite{Greshler2021a}.  Similarly, ~\cite{mizumoto2025lifelong} focus on a warehouse environment where transport robots and human workers must arrive at adjacent locations to collaboratively perform a picking task.  The package-exchange robot-routing problem (PERR) conceptualises cooperation as dynamic task re-allocation by allowing agents to swap payloads mid-journey~\cite{Ma2016a}. While valuable, these models are insufficient for our problem for two key reasons. First, they oversimplify the spatio-temporal nature of the cooperative act itself, treating it as an instantaneous event or an abstract trade-off rather than a sustained, physically coupled action. Moreover, these works typically focus exclusively on a single type of cooperative task, whereas our problem considers a more realistic scenario involving a mixture of both individual and multi-agent tasks.

Once a team of agents is formed, their joint movement can be modelled as a single, large entity. This is the discrete-space analogue to extensive research in continuous domains on \textit{formation control} and \textit{cooperative transportation}~\cite{Cortes2017}. In discrete MAPF, this subproblem is known as Multi-Agent Path Finding for Large Agents (LA-MAPF)~\cite{Li2019i}. In LA-MAPF, agents occupy multiple vertices, making conflict resolution more complex than in classical MAPF. Standard CBS is inefficient because its single-vertex constraints cannot resolve geometric collisions, motivating specialised solvers like Multi-Constraint CBS (MC-CBS)~\cite{Li2019i}. LA-MAPF thus provides the formalism for the execution phase of a cooperative task.

Our proposed CT-TAPF problem integrates task selection, team formation, and coordinated motion. Standard one-to-one TAPF solver like CBS-TA~\cite{Hoenig2018a} is ill-suited,  as their core assumption of assigning a unique agent to each task would necessitate an agent for every task \textit{slot}. This is both unrealistic in scenarios where agents are a scarce resource and computationally intractable, as it leads to a combinatorial explosion in team assignments. A more promising starting point is Task Conflict-Based Search (TCBS)~\cite{Henkel2019a}, which allows an agent to execute a sequence of tasks. However, naively applying its allocation philosophy—expanding a new branch for every agent-task combination—creates a different combinatorial explosion in our problem discussed in Section.~\ref{sec:task-expansion}. To address this, we propose an optimal solver CT-TCBS. Our core contribution is an incremental expansion strategy that manages team formation progressively, significantly improving success rates and addressing a critical gap in the literature.


\section{Problem Definition}
\label{sec:problemdef}

We address a multi-agent multi-task assignment and path-finding problem. The problem is defined by a set of core components.
The environment is an undirected 2-dimension graph $G=(V, E)$, where $V$ is a set of vertices and $E$ is a set of edges.
We consider a set of $n$ agents, $\mathcal{A}=\{a_1, ..., a_n\}$, each starting at an initial vertex $v_i^0 \in V$.
There is also a set of $m$ cooperative tasks, $\mathcal{T}=\{\tau_1, ..., \tau_m\}$.
The goal is to find a task-slot assignment and a set of conflict-free paths $\Pi = \{P_1, ..., P_n\}$ that collectively fulfill all tasks.
The optimal solution minimises SoC, the total operational time for all agents. 

A cooperative task $\tau_i \in \mathcal{T}$ requires $k_i$ agents to execute.
Each task is defined by a start configuration $V_i^S = \{v_{i,1}^S, ..., v_{i,k_i}^S\}$ and a goal configuration $V_i^G$, both of which are sets of $k_i$ distinct vertices that form a connected subgraph in $G$.
The vertices within $V_i^S$ are referred to as \textit{task slots}.
The assignment is slot-based: each agent $a_j$ is assigned to at most one unique slot $v_{i,s}^S$ for a single task $\tau_i$.
We assume the number of agents is sufficient for any single task, i.e., $n \ge \max_{i}(k_i)$.

The execution of a task $\tau_i$ is a dynamic, two-phase process.
First, in the \textit{assembly phase}, the $k_i$ agents assigned to the task travel independently from their current locations to their respective slots in $V_i^S$.
During this phase, the vertices of the start configuration $V_i^S$ are not reserved; other agents on different missions are free to pass through them.
This design, inspired by real-world logistics systems, keeps task locations unobstructed, generalising the framework for missions beyond warehousing, such as inspection or patrol~\cite{Portugal2011}.
The task materialises only at the moment the last of its assigned agents arrives and synchronisation is achieved.
At this point, the \textit{convoy phase} begins, and the team transforms into a single \textit{Convoy} (conceptually equivalent to a Large Agent in \cite{Li2019i}).
This Convoy is formally defined by a reference position $r$ and a rigid shape $\mathcal{S}_i$, representing the set of relative coordinate offsets of all member agents.
Consequently, the \textit{footprint} of an entity $E$ (whether a single agent or a Convoy) when anchored at location $u$ at timestep $t$ is defined as the set of occupied vertices $F(E, u, t) = \{ u + \delta \mid \delta \in \mathcal{S}_E , T=t\}$.
Upon reaching the goal configuration, the task is considered complete, the Convoy is immediately dissolved, and the participating agents are freed for subsequent tasks.

A valid solution must adhere to strict path and collision constraints.
The path for an agent $a_j$ is a time-indexed sequence of vertices, $P_j = (v_j^0, v_j^1, ..., v_j^{T_j})$, where the cost $|P_j|$ is defined as the total duration $T_j$, inherently capturing any time spent waiting for synchronisation.
For any time step $t$, a move from $v_j^t$ to $v_j^{t+1}$ is valid if the target vertex is either identical to or adjacent to the current one in the four cardinal directions.
Furthermore, the plan must be conflict-free.
Crucially, we enforce a geometric conflict definition to handle the spatial extent of cooperative teams.
A plan is valid if and only if it is free of generalised vertex conflicts.
Let $u$ and $v$ denote the locations of two entities $E_i$ and $E_j$ at timestep $t$, respectively.
A conflict occurs if their footprints overlap, i.e., $F(E_i, u, t) \cap F(E_j, v, t) \neq \emptyset$.
This definition generalises standard MAPF collisions, acknowledging that a conflict can occur even if the entities' reference positions are distinct ($u \neq v$).
In this paper, we only focus on vertex conflicts as modelling stricter generalised edge conflicts introduces significant computational overhead.

\section{CT-TCBS}
\label{sec:CT-TCBS}

\subsection{Main Concept}
\label{sec:CT-TCBS_overview}
To solve the CT-TAPF problem, we introduce Cooperative Transportation Task Conflict-Based Search (CT-TCBS), an optimal, two-level search algorithm. The high level performs an A* search on a constraint tree to explore the combinatorial space of task assignments, employing a prioritised expansion strategy that systematically resolves path conflicts before assigning agents to new or partially-formed tasks. At each node in the tree, a low-level planner computes optimal paths for both individual agents and multi-agent convoys. To efficiently resolve collisions involving these convoys, a challenge analogous to LA-MAPF, CT-TCBS adopts the MC-CBS framework, leveraging strategies such as \texttt{ASYM}, \texttt{SYM}, and \texttt{MAX-d}~\cite{Li2019i}. The specifics of the search loop, conflict resolution, and task expansion are detailed in the subsequent sections.

\begin{figure}[h!] 
\centering 
\begin{adjustbox}{max width=\linewidth}
\begin{tikzpicture}[
    parent anchor=south,
    child anchor=north,
    level distance=2.2cm,
    level 1/.style={sibling distance=2.9cm}, 
    level 2/.style={sibling distance=3.8cm},
    level 3/.style={sibling distance=3.8cm},
    level 4/.style={sibling distance=3.8cm},
    base_node/.style={draw, rectangle, rounded corners, align=center, font=\LARGE\bfseries}, 
    ta_node/.style={base_node, fill=blue!20},
    partial_node/.style={base_node, fill=orange!30},
    collision_node/.style={base_node, fill=red!20},
    final_node/.style={base_node, fill=green!20},
    narrow/.style={minimum width=2.2cm, minimum height=1.5cm},
    wide/.style={minimum width=3cm, minimum height=1.5cm},
    edge_label/.style={font=\small\bfseries, inner sep=1pt, fill=white, circle} 
]

\node[ta_node] {
    \textbf{1} \\
    TA: \{\{\},\{\},\{\}\} \\
    CS: \{\{\},\{\},\{\}\}
}
    child { node[ta_node] {\textbf{2} \\ TA: \{\{1\},\{\},\{\}\} \\ CS: \{\{\},\{\},\{\}\}} edge from parent node[below, pos=0.3, edge_label] {1} }
    child { node[ta_node] {\textbf{3} \\ TA: \{\{\},\{1\},\{\}\} \\ CS: \{\{\},\{\},\{\}\}} edge from parent node[below, pos=0.3, edge_label] {1} }
    child { node[ta_node] {\textbf{4} \\ TA: \{\{\},\{\},\{1\}\} \\ CS: \{\{\},\{\},\{\}\}} edge from parent node[below, pos=0.3, edge_label] {1} }
    child { node[partial_node] {
        \textbf{5} \\
        TA: \{\{2.0\},\{\},\{\}\} \\
        CS: \{\{\},\{\},\{\}\}
    }
        child { node[collision_node] {
            \textbf{8} \\
            TA: \{\{2.0\},\{2.1\},\{\}\} \\
            CS: \{{\textcolor{red}{Collision}}\}
        }
            child { node[ta_node] {
                \textbf{10} \\
                TA: \{\{2.0\},\{2.1\},\{\}\} \\
                CS: \{\{((3, 2), 5)\},\{\},\{\}\}
            }
                child { node[final_node] {\textbf{12} \\ TA: \{\{2.0,1\},\{2.1\},\{\}\} \\ CS: \{\{((3, 2), 5)\},\{\},\{\}\}} edge from parent node[below, pos=0.3, edge_label] {4} }
                child { node[final_node] {\textbf{13} \\ TA: \{\{2.0\},\{2.1,1\},\{\}\} \\ CS: \{\{((3, 2), 5)\},\{\},\{\}\}} edge from parent node[below, pos=0.3, edge_label] {4} }
                child { node[final_node, draw=blue, line width=2pt] {\textbf{14} \\ TA: \{\{2.0\},\{2.1\},\{1\}\} \\ CS: \{\{((3, 2), 5)\},\{\},\{\}\}} edge from parent node[below, pos=0.3, edge_label] {4} }
            edge from parent node[below, pos=0.3, edge_label] {3}
            }
            child { node[ta_node] {
                \textbf{11} \\
                TA: \{\{2.0\},\{2.1\},\{\}\} \\
                CS: \{\{\},\{((3, 2), 5)\},\{\}\}
            } edge from parent node[below, pos=0.3, edge_label] {3} }
        edge from parent node[below, pos=0.3, edge_label] {2}
        }
        child { node[ta_node] {
            \textbf{9} \\
            TA: \{\{2.0\},\{\},\{2.1\}\} \\
            CS: \{\{\},\{\},\{\}\}
        } edge from parent node[below, pos=0.3, edge_label] {2} }
    edge from parent node[below, pos=0.3, edge_label] {1}
    }
    child { node[partial_node] {\textbf{6} \\ TA: \{\{\},\{2.0\},\{\}\} \\ CS: \{\{\},\{\},\{\}\}} edge from parent node[below, pos=0.3, edge_label] {1} }
    child { node[partial_node] {\textbf{7} \\ TA: \{\{\},\{\},\{2.0\}\} \\ CS: \{\{\},\{\},\{\}\}} edge from parent node[below, pos=0.3, edge_label] {1} };
\end{tikzpicture}
\end{adjustbox}
\caption{The incremental search tree of CT-TCBS for an instance with 3 agents and 2 tasks. Colours indicate node state: blue for standard, orange for partially assigned, and red for collision.}
\label{fig:search_tree} 
\Description{A search tree diagram illustrating the incremental expansion process of the CT-TCBS algorithm. The root node is at the top, branching down into child nodes representing different states of task allocation and constraints. Nodes are colour-coded: blue for standard, orange for partially assigned tasks, and red for states with collisions. Edges are numbered to show the best-first search order, culminating in a final solution node.}
\end{figure}
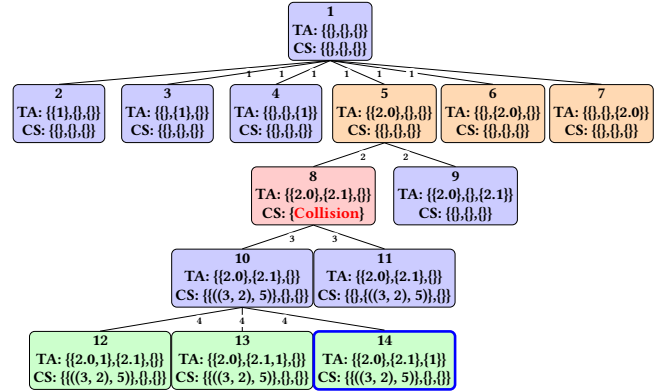

An example of the incremental search process is shown in Figure.~\ref{fig:search_tree}. The numbers within each node and on the connecting edges denote the node's unique ID and the expansion step order. \textbf{TA} denotes Task Allocation and \textbf{CS} denotes the Constraint Set, with each pair of curly braces corresponding to one of the three agents in the problem. For instance, the state `TA: \{\{2.0\},\{\},\{\}\}' in node 5 indicates the first agent is assigned to slot 2.0 of the two-agent task 2. Similarly, `CS: \{\{((3, 2), 5)\},\{\},\{\}\}' in node 10 shows a constraint added to the first agent, prohibiting it from location (3,2) at timestep 5.

The search initiates at the root (Node 1) with an empty plan, expanding into a layer of nodes, each with one assigned \textit{task slot}. Then, the lowest f-cost node 5 is chosen for expansion. This leads to the generation of node 8, where a collision is detected. Because conflict resolution is prioritised, node 8 is expanded next in step 3. The search continues this process until a complete and conflict-free solution, represented by node 14, is found.

\subsection{High-Level Search}
The main loop of CT-TCBS operates as a best-first search, guided by the A* tree search. This search explores a constraint tree where each node represents a partial solution, defined by a specific set of task assignments and spatio-temporal constraints that agents must adhere to. At the core of this search is the evaluation of each node $N$ using the standard A* cost function: $f(N) = g(N) + h(N)$. Here, $g(N)$ represents the \textit{cost-so-far}: the true optimal cost of the plan for all tasks that have been assigned within node $N$, respecting its current constraints. This value is computed by the low-level planner. The second component, $h(N)$, is the \textit{heuristic cost}: an admissible estimate of the minimum additional cost required to complete all currently unassigned tasks.

The search process begins by initialising an \textit{OPEN list} with a root node, which contains empty task assignments and no constraints. In each step of the main loop, the algorithm selects the node with the lowest f-value from the OPEN list for expansion.

Upon selecting a node, the algorithm first performs a goal test. A node is considered a goal, and thus represents an optimal solution, if two conditions are met: (1) all tasks in the problem instance have been fully assigned to agent teams, and (2) the corresponding path plan generated by the low-level planner is conflict-free. If the node fails this test, it is expanded to generate successor states. The expansion logic follows a strict priority order, which is detailed in Sections~\ref{sec:conflict-expansion} and ~\ref{sec:task-expansion}, where resolving existing conflicts is always prioritised over assigning new tasks.

During implementation, to ensure search efficiency, we employ a CLOSED list to handle duplicate states. As the incremental assignment process can generate nodes with identical task assignments and constraints via different branches, this list prevents the redundant exploration of previously expanded states.

\begin{algorithm}[H]
\caption{CT-TCBS High-Level Search}
\label{algo:main_loop}
\begin{algorithmic}[1]
\Require Initial state of agents, a set of all tasks
\Ensure The optimal solution or FAILURE

\State $root \gets \text{new Node()}$ \Comment{1. Initialisation}
\State $root.g_{cost} \gets 0$
\State $root.h_{cost} \gets \text{Heuristic}(root)$
\State $OPEN.push(root)$
\State $CLOSED \gets \emptyset$
\While{$OPEN$ is not empty} \Comment{2. Main Search Loop}
    \State $currentNode \gets OPEN.pop()$
    \If{\text{Goal\_Test}($currentNode$)} \Comment{3. Goal Test}
        \State \Return Construct\_Solution($currentNode$) 
    \EndIf
    \State $successorStates \gets \text{Expand}(currentNode)$
    \For{each $state$ in $successorStates$} 
        \If{$state \in CLOSED$}
        \State \textbf{continue}
        \EndIf
        \State $CLOSED.add(state)$
        \State $child \gets new Node(state)$
        \State $child.paths, child.g_{cost}, child.conflict \gets$ 
        \Statex \quad\quad\quad $\text{Cost\_So\_Far}(child)$ \Comment{4. Cost so far}
        \If{$child.paths$ is not NULL} 
            \State $child.h_{cost} \gets \text{Heuristic}(child)$ \Comment{5. Heuristic}
            \State $OPEN.push(child)$
        \EndIf
    \EndFor
\EndWhile

\State \Return \text{FAILURE}
\end{algorithmic}
\end{algorithm}

\subsection{Low-Level Pathfinding}
\label{sec:low_level}

We employ an iterative, state-aware planner to translate abstract assignments into concrete trajectories.
A \textit{Unified A*} search is used to compute optimal paths respecting spatio-temporal constraints for both individual agents and multi-agent Convoys.
For fully assigned cooperative tasks, planning follows a three-step synchronisation:
(1) it computes individual paths to assembly slots to determine arrival times;
(2) it calculates the synchronisation time $t_{sync} = \max (t_{arr}^j)$;
and (3) it plans the team's joint movement as a single Convoy starting at $t_{sync}$.
For partially assigned teams in the search tree, paths are computed only up to the assembly slots to ensure the node's $g$-cost is admissible and informative.

\subsection{Conflict Expansion}
\label{sec:conflict-expansion}

When the low-level planner detects a collision in a node's current plan, the Conflict Expansion procedure is invoked. A conflict can occur either as an agent travels from its previous location to a task's start position, or during the joint convoy execution of a task. This leads to three potential conflict types: a collision between two single agents, a single agent and a convoy, or two convoys. For collisions involving convoys (i.e., large agents), the \texttt{Normal} Conflict-Based Search (CBS) approach of adding a single constraint to each branch is known to be inefficient~\cite{Li2019i}. To address this, we adopt the MC-CBS framework, which resolves a geometric collision by adding multiple constraints at once. Specifically, we consider three distinct MC-CBS strategies: Asymmetric (\texttt{ASYM}), Symmetric (\texttt{SYM}), and MaxWeight-d (\texttt{MAX-d})~\cite{Li2019i}.

\texttt{ASYM} resolves a vertex conflict $\langle a_i, a_j, u, v, t \rangle$ by creating an unbalanced split. One child node receives a single-constraint set, formally $\{\langle a_i, u, t \rangle\}$, which prohibits agent $a_i$ from being at its conflicting vertex. The other child node receives a large constraint set, defined as $\{\langle a_j, v', t \rangle \mid \langle a_i, a_j, u, v', t \rangle\}$ is a vertex conflict, which prohibits agent $a_j$ from being at any vertex $v'$ where it could collide with $a_i$ (if $a_i$ remains at vertex $u$) at timestep $t$.

In contrast, \texttt{SYM} creates a more balanced resolution. It chooses a point $p$ in the Euclidean space that is inside the geometric overlap area of the two agents, $S_i(u) \cap S_j(v)$. It then adds one constraint set to each child node, where each set blocks all vertices that would cause the agent's shape to include point $p$. These sets are formally defined as $C_1 = \{\langle a_i, v', t\rangle \mid p \in S_i(v'), v' \in V\}$ and $C_2 = \{\langle a_j, v', t\rangle \mid p \in S_j(v'), v' \in V\}$. Unlike \texttt{ASYM}, this method typically results in child nodes receiving constraint sets of more comparable sizes.

\texttt{MAX-d} offers a more sophisticated strategy for selecting constraints. The goal of \texttt{MAX-d} is to make the most progress in the high-level search by choosing constraints that maximally increase the costs of the child nodes. To achieve this, it uses a Multi-Valued Decision Diagram (MDD) with a lookahead depth of $d$ to predict the cost increase (the `weight') that a set of constraints will impose on an agent's path. For a given conflict, \texttt{MAX-d} analyses potential constraint sets for both agents and selects the pair that is mutually disjunctive and maximises the smaller of the two weights, effectively prioritising the most impactful or cardinal conflicts.

While \texttt{MAX-d} is documented to have the best performance for the LA-MAPF subproblem, we hypothesise that its effectiveness may be reduced in our broader CT-TAPF setting. The primary search challenge in CT-TAPF is often the enormous combinatorial space of task allocation, rather than pathfinding conflicts. The strategy of \texttt{MAX-d}, which deliberately increases the g-cost to prune the conflict search tree, may force the high-level search to expand more nodes in the task allocation space to find a new, cheaper assignment. 

Finally, it is essential that conflict expansion is prioritised over any form of task expansion. If a node with unresolved conflicts were to be expanded for task assignment, its f-value would be an overly optimistic underestimation of the true cost to reach a solution through that path. By resolving all known conflicts first, we ensure the integrity of the node costs, which is fundamental to the efficiency and optimality of the A* search.

\subsection{Task Expansion}
\label{sec:task-expansion}

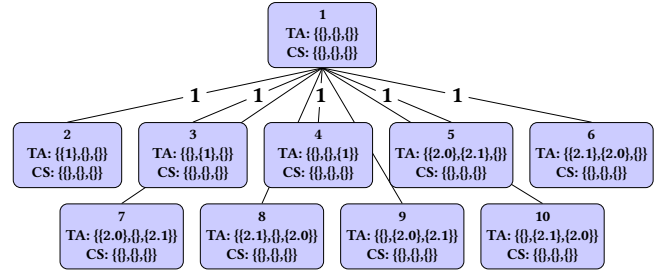
\begin{figure}[h!]
\centering
\usetikzlibrary{positioning} 

\pgfdeclarelayer{background}
\pgfdeclarelayer{main}
\pgfsetlayers{background,main} 

\begin{adjustbox}{max width=\linewidth}
\begin{tikzpicture}[
    base_node/.style={draw, rectangle, rounded corners, align=center, font=\small\bfseries},
    ta_node/.style={base_node, fill=blue!20, minimum height=1.2cm, minimum width=2.0cm},
    edge_label/.style={font=\LARGE\bfseries, inner sep=1pt, fill=white, circle, midway}
]

\begin{pgfonlayer}{main}
    \node[ta_node] (root) {
        \textbf{1} \\
        TA: \{\{\},\{\},\{\}\} \\
        CS: \{\{\},\{\},\{\}\}
    };

    \node[ta_node, below=1.0cm of root, xshift=-4.7cm] (c2) {\textbf{2} \\ TA: \{\{1\},\{\},\{\}\} \\ CS: \{\{\},\{\},\{\}\}};
    \node[ta_node, right=0.3cm of c2] (c3) {\textbf{3} \\ TA: \{\{\},\{1\},\{\}\} \\ CS: \{\{\},\{\},\{\}\}};
    \node[ta_node, right=0.3cm of c3] (c4) {\textbf{4} \\ TA: \{\{\},\{\},\{1\}\} \\ CS: \{\{\},\{\},\{\}\}};
    \node[ta_node, right=0.3cm of c4] (c5) {\textbf{5} \\ TA: \{\{2.0\},\{2.1\},\{\}\} \\ CS: \{\{\},\{\},\{\}\}};
    \node[ta_node, right=0.3cm of c5] (c6) {\textbf{6} \\ TA: \{\{2.1\},\{2.0\},\{\}\} \\ CS: \{\{\},\{\},\{\}\}};

    \node[ta_node, below=2.5cm of root, xshift=-3.7cm] (c7) {\textbf{7} \\ TA: \{\{2.0\},\{\},\{2.1\}\} \\ CS: \{\{\},\{\},\{\}\}};
    \node[ta_node, right=0.3cm of c7] (c8) {\textbf{8} \\ TA: \{\{2.1\},\{\},\{2.0\}\} \\ CS: \{\{\},\{\},\{\}\}};
    \node[ta_node, right=0.3cm of c8] (c9) {\textbf{9} \\ TA: \{\{\},\{2.0\},\{2.1\}\} \\ CS: \{\{\},\{\},\{\}\}};
    \node[ta_node, right=0.3cm of c9] (c10) {\textbf{10} \\ TA: \{\{\},\{2.1\},\{2.0\}\} \\ CS: \{\{\},\{\},\{\}\}};
\end{pgfonlayer}

\begin{pgfonlayer}{background}
    \foreach \i in {c2, c3, c4, c5, c6, c7, c8, c9, c10} {
        \path (root.south) edge node[edge_label] {1} (\i.north);
    }
\end{pgfonlayer}

\end{tikzpicture}
\end{adjustbox}
\caption{An illustration of the Combinatorial explosion caused by the \textit{Combinatorial Expansion} strategy. For a 3-agent problem, assigning a two-agent task instantly generates all 6 possible team permutations (nodes 5-10) in a single step from the root, creating a large branching factor which the incremental approach (Figure.~\ref{fig:search_tree}) avoids.}
\Description{A search tree diagram showing the combinatorial explosion of the naive Combinatorial Expansion strategy. A single root node at the top expands into ten child nodes in a single step. This illustrates a large branching factor, contrasting with the more controlled incremental approach shown in the previous figure.}
\label{fig:combinatorial_tree_2rows_layered}
\end{figure}

Task Expansion is performed to explore new task assignments, if a node is conflict-free but does not satisfy the goal condition. A naive approach, which we term \textit{Combinatorial Expansion}, would be to generate a child node for every possible assignment of an unassigned task. For a multi-agent task requiring $k$ agents, this involves considering every possible coalition of $k$ available agents, creating a new child node for each valid team. An example of this expansive branching is shown in Figure.~\ref{fig:combinatorial_tree_2rows_layered}. This method leads to a combinatorial explosion in the branching factor, making it computationally intractable for all but the simplest of problem instances.

To overcome this challenge, we introduce our primary contribution for the high-level search: \textit{Incremental Expansion}. Instead of assigning a full team at once, this strategy dramatically reduces the branching factor by breaking the assignment process into a sequence of prioritised steps. The logic is as follows:
\begin{enumerate}
    \item If a multi-agent task is already partially assigned, the algorithm performs a \textbf{Partial Task Completion Expansion}, creating a new child node for each available agent that can fill the next open slot of that task.
    \item Only when all tasks are fully assigned, does the algorithm perform an \textbf{Assign New Task Expansion}. In this step, it selects a new, unassigned task and creates child nodes by assigning a single available agent to its first slot.
\end{enumerate}
This creates a series of intermediate, `not fully executable' nodes (visualised by their orange colour in Figure.~\ref{fig:search_tree}) and ensures a much smaller branching factor at each step. For our empirical analysis, we also developed a hybrid variant, \textit{Incremental with Large Root (Incremental-LR)}, which slightly increases the initial branching for a new task by considering all its $k$ slots, not just the first one. Our hypothesis is larger branch factor results in worse performance.

The introduction of `not fully executable' nodes necessitates a modification to how the \textit{cost-so-far}, $g(N)$, is calculated. For any agent assigned to a task that is still incomplete (i.e., not all slots are filled), its determined cost is only the cost of its path from its last known location to its designated slot for the task. The costs for waiting for its collaborators to arrive and for the joint execution of the task are not yet determined and are therefore left to be estimated by the heuristic function.

This incremental approach reveals a fundamental trade-off, which we call the \textit{task-conflict expansion dilemma}. The combinatorial strategy creates a search tree that is extremely wide in terms of task assignments but relatively shallow in terms of conflict resolution depth, as a complete team is evaluated at once. In contrast, our incremental strategy creates a much narrower tree at each task-assignment step, but it may require more subsequent conflict-resolution steps as the team is built piece by piece. However, we posit that in the CT-TAPF problem, the search space for task allocation is the dominant factor. Therefore, although the dilemma exists, the incremental method's significant reduction of the task-assignment branching factor makes it a more efficient and scalable strategy.

\subsection{Heuristic Function}
\label{sec:heuristic}
To guide the A* search efficiently, we introduce a composite heuristic function $H$.
For the search to guarantee optimality, $H$ must be \textbf{admissible}, meaning $H(n) \leq h^*(n)$ for any node $n$, where $h^*(n)$ is the true minimal cost to reach the goal. An ideal heuristic would estimate all future costs, which can be divided into three distinct, non-overlapping components: the transport cost for partially-assigned tasks ($H_1$), the assignment and transport cost for all unassigned tasks ($H_2$), and the team synchronisation waiting time ($H_3$).
Our final heuristic is constructed by summing the admissible estimators for the computationally tractable components.

$H_1$ estimates the committed future transport cost. This component accounts for the certain future costs for agents that are already assigned to a multi-agent task that is not yet fully staffed. For such an agent, the cost of executing its part of the task—travelling from the task's start position to its goal position—is guaranteed to be incurred. We calculate a lower bound for this cost using the Manhattan distance, which provides a simple and admissible estimate, as the true path cost will always be greater than or equal to this distance.

$H_2$ estimates the future assignment cost for all task slots that are currently unassigned. The precise method for ensuring its admissibility is critically dependent on the overall optimisation objective. For our SoC objective, which minimises the total travel distance, the heuristic must account for the two fundamental ways a slot might be filled in the optimal plan: an agent may travel directly from its current location, or it may first complete another unassigned task and then proceed to the current task's start in a chained assignment. To create a valid lower bound, our heuristic therefore calculates the cost for both possibilities—the minimal cost from any available agent to the slot (for the direct scenario) and the minimal cost from any other task's goal to the slot (for the chained scenario)—and takes the minimum. 

A third potential component, $H_3$, would estimate the future team synchronisation wait time, a cost unique to multi-agent tasks. While an accurate $H_3$ could significantly improve search performance, calculating a non-trivial, admissible lower bound is computationally intractable under a SoC objective. To do so would require solving a subproblem equivalent to the Generalised Assignment Problem (GAP)~\cite{Sahni1976} to find the optimal conflict-free team of agents and potential intermediate tasks—a problem known to be NP-hard. Given this inherent complexity, we exclude this component from our heuristic.

In summary, our total heuristic is the sum of the two tractable components: $H = H_1 + H_2$. Since $H_1$ and $H_2$ are admissible estimators for distinct phases of task completion (committed transport vs. future unassigned tasks), their sum remains a guaranteed admissible heuristic for the total remaining cost. This admissibility is a cornerstone of our algorithm's optimality.

\subsection{Optimality of CT-TCBS}
\label{sec:optimality_CT-TCBS}

The optimality of our CT-TCBS algorithm is rooted in its foundation as an A* search. To guarantee that an optimal solution is found, several properties must hold: the heuristic function must be admissible, all step costs in the search tree must be non-negative, and the expansion mechanisms must be sound.

The first property is satisfied by our design. As established in Section~\ref{sec:heuristic}, the composite heuristic, $H$, is the sum of three distinct and individually admissible components, guaranteeing that the overall heuristic never overestimates the true cost to a goal state.

For the second property, the cost $g(N)$ of any node $N$ is the optimal value of a combinatorial optimisation problem defined by a set of task assignments and spatio-temporal constraints. A child node $N_c$ is generated from a parent $N_p$ by adding a new constraint (either a new task assignment or a path constraint). Consequently, the set of all valid, conflict-free plans that satisfy the child's constraints is a subset of the valid plans for the parent. The relaxation principle indicates that minimising an objective function over a subset of a feasible domain cannot yield a value lower than minimising it over the full domain. Thus, it is guaranteed that $g(N_c) \ge g(N_p)$, and the non-negative step cost condition holds.

Finally, the conflict resolution mechanism does not compromise optimality. Our framework employs MC-CBS methods, which function by adding valid constraints to the search. This process correctly prunes branches of the search tree that contain collisions but is guaranteed never to prune the optimal, collision-free solution path, a principle well-established in the CBS literature.

\subsection{Suboptimal Variant: Task Selector Layer BT \& WT}
\label{subsec: subopt BT WT}

While optimal algorithms provide a crucial theoretical benchmark, their computational cost often makes them intractable for larger problem instances. To balance solution quality with runtime, we introduce a family of suboptimal solvers. A natural starting point, inspired by the TCBS, is the agent-centric nearest neighbour \texttt{`-nn'} family of algorithms. It prunes the search tree by restricting each agent to consider only the `n' tasks in its closest proximity, sacrificing guaranteed optimality for a smaller search space~\cite{Henkel2019a}.

However, this agent-centric, proximity-based heuristic proves fundamentally ill-suited for the cooperative demands of the CT-TAPF problem. An agent myopically selecting its closest task may commit to a multi-agent task whose other required slots are far from any available partners. This can lead to globally inefficient solutions with excessive waiting times or, in worse cases, deadlocks where partially assigned teams can never be completed. The core issue is that a locally optimal choice for one agent ignores the global logistics of team formation.

To address the limitations of this myopic, agent-centric view, we propose a novel suboptimal strategy that adopts a global, \textbf{task-centric} perspective. Instead of allowing each agent to choose from a local set of tasks, our algorithm inserts a task selection layer that globally decides which single task is the most strategic to assign next. This layer prunes the search space by expanding only one chosen task, which can either be the \textit{Best Task (BT)}, the easiest to complete, or the \textit{Worst Task (WT)}, the most difficult.

We formalise \textit{task difficulty} as the minimum estimated cost to complete a given task with the currently available agents. To compute this efficiently, we construct a cost matrix where each entry represents the sum of an agent's unconstrained travel time to a task slot plus the task's execution time. The optimal assignment of a full team and its corresponding minimum cost is then found by solving this assignment problem using the Hungarian algorithm. Since the task execution time is constant for any agent assigned to that task, the algorithm inherently finds the team with the minimum possible sum of arrival times. This sum serves as an indicator for the overall team synchronisation cost, making a separate, more complex waiting time calculation unnecessary for this heuristic selector.

The rationale for these two opposing strategies, BT and WT, is rooted in established principles of heuristic search. Selecting the best (lowest-cost) task is a \textit{greedy} approach that aims to find a high-quality solution quickly by prioritising the easiest parts of the problem. Conversely, selecting the worst (highest-cost) task is a \textit{fail-fast} strategy, common in constraint satisfaction problems. This approach tackles the most constrained or difficult parts of the problem first, with the goal of pruning large, unviable branches of the search tree early.


\section{Experiment}
\label{sec:experiment}


We empirically evaluate our proposed algorithms on a variety of CT-TAPF instances generated on grid maps. Each instance is characterised by the number of agents $n$ with their initial positions, and a set of tasks, where each task $\tau_i$ requires $k_i$ agents (where $k_i$ is between 1 and 4) and is defined by a start and a goal configuration. To test algorithm performance under diverse conditions, we introduce three distinct scenarios. Our baseline is a \textbf{random} scenario set, where agents and tasks are placed uniformly at random on any vacant vertex. To simulate high-traffic conditions, we developed a \textbf{spatially-biased} scenario set. This generator utilises probabilistic heatmaps with opposing linear gradients to concentrate task endpoints in opposite corners, thereby forming a congested central corridor and biasing agent start positions towards the remaining corners to ensure maximum travel distances. Finally, to specifically evaluate conflict resolvers, we designed a \textbf{collision-rich} scenario set using a placement heuristic. Once an initial path is established, this method places subsequent task endpoints on opposite sides of its bounding box to ensure trajectory intersection, thereby creating instances with a high likelihood of inter-convoy conflicts.

We evaluate our proposed algorithms, three optimal CT-TCBS variants and the suboptimal CT-TCBS-BT/WT solvers, against several baselines that represent different points on the solution quality-runtime trade-off spectrum. These baselines include two agent-centric methods from prior work, \texttt{-nn1} and \texttt{-nn2}, and a priority-based heuristic, \textit{Greedy-PP}, which we adapt from~\cite{Henkel2019a}. The adapted \texttt{Greedy-PP} baseline is an iterative solver that prevents deadlocks under cooperative context by greedily selecting the easiest task, determined by a minimum-cost Hungarian assignment of agents to task slots, and then sequentially planning its complete, conflict-free path using a reservation table before considering the next task. The evaluation is structured into two distinct experimental sets. The optimal solvers were analysed on 25 instances generated on an 8x8 empty grid using the \textbf{collision-rich} scenario, with scenarios scaling up to \emph{8 tasks and 6 agents}, where all tasks exclusively required a team of two. Subsequently, to facilitate a holistic comparison across all methods, we evaluated both optimal and suboptimal algorithms on a separate set of 50 instances on a 16x16 grid with 10\% obstacle density, comprising 25 from the \textbf{random} scenario and 25 from the \textbf{spatially-biased} scenario. These larger-scale tests scaled up to \emph{15 tasks and 5 agents}, featuring a diverse mix of cooperative requirements; for instance, the largest scenario instance included nine 1-agent tasks, three 2-agent tasks, two 3-agent tasks, and one 4-agent task. All algorithms were implemented in Python.

To ensure a fair comparison, our testing pipeline is inspired by the MovingAI Lab's MAPF benchmark~\cite{Sturtevant2012a}. We solve instances incrementally, starting with a small problem size and only proceeding to a larger one if the current instance is solved within a 500-second timeout and a 4GB memory limit. The problem size is scaled using two parameters: a \textit{\textbf{fixed task-type ratio}} dictating the proportion of tasks requiring different numbers of agents, and a \textit{\textbf{task-agent ratio}} setting the number of agents relative to the total task slots. To prevent large jumps in difficulty, we employ a weighted round-robin algorithm. This method adds only one task at each incremental step, selecting the task type that best maintains the overall distribution close to the target task-type ratio. All experiments were distributively executed on two 48-core Intel(R) Xeon(R) Cooper Lake @ 3.30 Ghz servers running Ubuntu 24.04. Our source code and datasets are available at \url{https://github.com/BlankNing/CTAPF_CT_TCBS}.


\begin{figure}[h]
  \centering
 \includegraphics[width=1\linewidth]{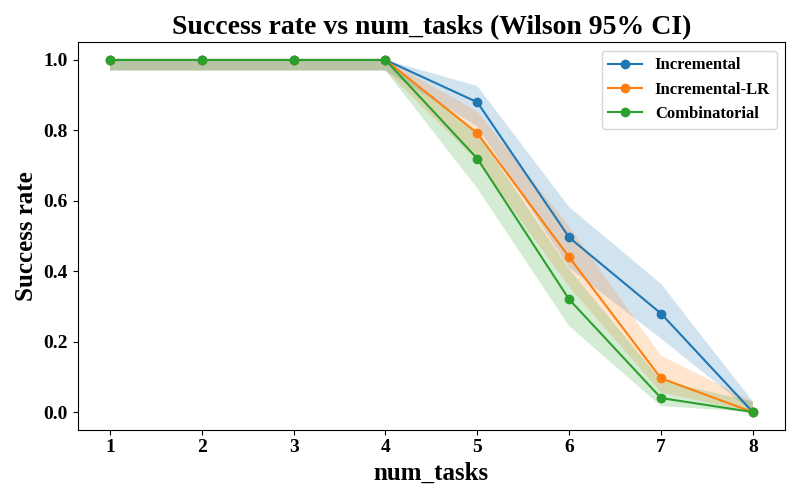}
  \caption{Success rates on collision-rich set with different optimal expansion strategies.}
  \label{fig:success_rate_collision_expansion}
  \Description{A line graph comparing the success rate of three optimal expansion strategies: Incremental, Incremental-LR, and Combinatorial. The x-axis represents the number of tasks, from 1 to 8. The y-axis represents the success rate, from 0.0 to 1.0. All three strategies show a decreasing success rate as the number of tasks increases, with the `Incremental' strategy consistently maintaining the highest success rate.}
\end{figure}

\subsection{Optimal Analysis}

We first evaluated the performance of the three optimal expansion strategies on the \textbf{collision-rich} scenario set. As hypothesised and shown in Figure.~\ref{fig:success_rate_collision_expansion}, the \texttt{Incremental} strategy consistently outperforms both \texttt{Incremental-LR} and \texttt{Combinatorial}. This observation is statistically significant: a Cochran's Q test confirmed a significant difference among the strategies ($W_Q\approx0.053,\ p \approx 10^{-5}$), and post-hoc paired McNemar's tests confirmed that \texttt{Incremental}'s success rate is significantly higher than both \texttt{Incremental-LR} ($g = 1.0,\ p_{Holm} \approx 0.047$) and \texttt{Combinatorial} ($g=1.0,\ p_{Holm} \approx 0.001$). 

To diagnose this gap we analysed search behaviour. Table~\ref{tab:avg_tecerr_rank_optimal_collision} shows a systematic trade-off, which is proven to be statistically significant using Friedman test with post-hoc Paired Wilcoxon signed-rank test ($W\approx0.493,\ p \approx 10^{-28}$ \text{for task};\ $W\approx0.121,\ p \approx 10^{-7}$ \text{for conflict}). \texttt{Combinatorial} path conflicts with the lowest average rank, but it also has the highest average task-expansion rank. In contrast, \texttt{Incremental} yields best average task-expansion rank, while \texttt{Incremental-LR} has the worst rank for conflict expansions. Quantitatively, across all successfully solved instances, task-expansion nodes substantially dominate conflict-expansion nodes (median ratio $\tilde{r}\!=\!9.54$, IQR $[5.64,\,20.0]$) among cases with nonzero conflict expansions, indicating that the combinatorial burden of \emph{task allocation—not pathfinding conflicts—is the primary computational bottleneck.} The superior performance of \texttt{Incremental} therefore stems from more effective pruning of this dominant portion of the search space.

\begin{table}[t]
	\caption{Average ranks for task and conflict expansions among optimal expansion strategies on \textbf{collision-rich} set (lower is better).}
	\label{tab:avg_tecerr_rank_optimal_collision}
	\begin{tabular}{rll}\toprule
		\textit{Expansion} & \textit{Task Avg Rank} & \textit{Conflict Avg Rank} \\ \midrule
		Combinatorial & 2.250 & \textbf{1.217} \\
		Incremental-LR & 2.184 & 1.724  \\
		Incremental & \textbf{1.289} & 1.539 \\ 
        \bottomrule
	\end{tabular}
\end{table}


 We then compared five different conflict resolvers on the same \textbf{collision-rich} scenario set, as visualised in Table~\ref{tab:tece_resolver_rankshare_combined}. While the final success rates revealed no statistically significant difference among the resolvers, we observed a consistent, albeit slight, trend where the sophisticated \texttt{MAX-d} variants performed worse than other resolvers. This observation motivated us to investigate the underlying search behaviour to uncover potential efficiency differences. A Friedman test confirmed that the choice of resolver has a statistically significant impact on both task ($W\approx0.10,\ p \approx 10^{-10}$) and conflict ($W\approx0.130,\ p \approx 10^{-15}$) expansions. Post-hoc tests verified that the \texttt{MAX-d} variants expand significantly more nodes than other methods. Statistical data can be accessed in Appendix. 

We conjecture that this is a direct consequence of the `Task-Conflict Dilemma': \texttt{MAX-d}'s core strategy of deliberately increasing a node's g-value to prune the conflict-resolution tree has a negative side-effect in the integrated CT-TAPF problem. The inflated g-value makes the current task assignment appear more costly to the high-level A* search, \emph{prematurely forcing the algorithm to explore alternative task allocations}. This leads to a larger overall search and, consequently, a lower success rate within the time limit.

\begin{table}[t]
\centering
\small
\caption{Resolver comparison on the \textbf{collision-rich} set. Values are \emph{mean ranks} (smaller is better). Column headers: \texttt{Inc.} (Incremental), \texttt{Inc-LR} (Incremental-LR), and \texttt{Comb.} (Combinatorial). The \texttt{MAX-d} variants consistently rank worst (bold).}
\label{tab:tece_resolver_rankshare_combined}
\begin{tabular}{lcccccc}
\toprule
\multirow{2}{*}{Resolver} & \multicolumn{3}{c}{\textbf{Task Expansions}} & \multicolumn{3}{c}{\textbf{Conflict Expansions}} \\
\cmidrule(lr){2-4} \cmidrule(lr){5-7}
& Inc. & Inc-LR & Comb. & Inc. & Inc-LR & Comb. \\
\midrule
\texttt{ASYM}   & 1.114          & 1.153          & 1.079          & 1.151          & 1.204          & 1.112 \\
\texttt{SYM}    & 1.078          & 1.115          & 1.026          & 1.054          & 1.051          & 1.039 \\
\texttt{Normal} & 1.139          & 1.070          & 1.053          & 1.301          & 1.236          & 1.164 \\
\texttt{MAX-1}  & 1.428          & 1.427          & 1.355          & \textbf{1.614} & 1.618          & \textbf{1.546} \\
\texttt{MAX-2}  & \textbf{1.446} & \textbf{1.478} & \textbf{1.375} & 1.590          & \textbf{1.637} & 1.520 \\
\bottomrule
\end{tabular}
\end{table}

\subsection{Suboptimal Analysis}

\begin{figure}[h]
  \centering
    \includegraphics[width=1\linewidth]{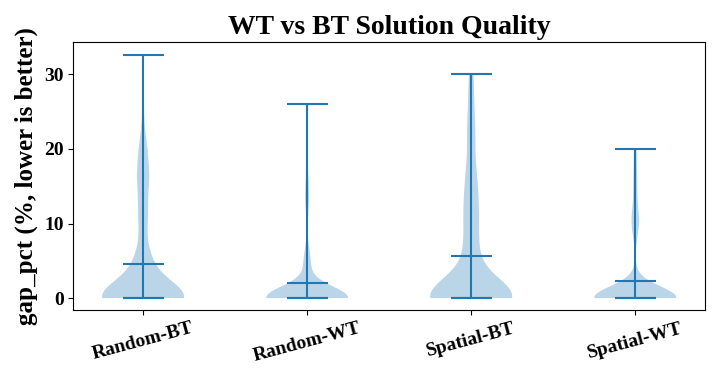}
\caption{Solution quality (optimality gap, \%) of WT vs.\ BT on Random and Spatially-biased scenario sets under sparse-10 maps.
WT achieves substantially lower means and variances than BT in both sets, indicating better solution quality.}
\label{fig:subopt_gap_selector_random}
  \Description{Four violin plots comparing the solution quality (optimality gap percentage) of the WT and BT selectors. The plots are grouped into two pairs: Random-BT vs. Random-WT, and Spatial-BT vs. Spatial-WT. The y-axis represents the gap percentage. In both pairs, the WT selector's violin plot is shorter and wider at the bottom, indicating a lower mean and smaller variance in the optimality gap compared to the BT selector.}
\end{figure}

For our suboptimal analysis, we evaluated our task-centric selectors, \texttt{Worst-Task (WT)} and \texttt{Best-Task (BT)}. The results reveal a clear and statistically significant hierarchy in solution quality. A paired Wilcoxon signed-rank test confirmed that the \texttt{WT} selector produces solutions with a significantly smaller gap to the optimal cost compared to \texttt{BT} on both random and spatially-constrained sets, as shown in Figure.~\ref{fig:subopt_gap_selector_random} ($r=0.294,\ p \approx 10^{-11}$). We attribute this to \texttt{WT}'s strategy of prioritising large, multi-agent tasks. This forms the necessary agent teams early, after which agents can efficiently complete simpler tasks. In contrast, \texttt{BT} often leaves agents scattered after completing simple tasks, forcing them to incur significant travel costs to rendezvous for cooperative tasks later.

\begin{figure}[h]
  \centering
  \includegraphics[width=1\linewidth]{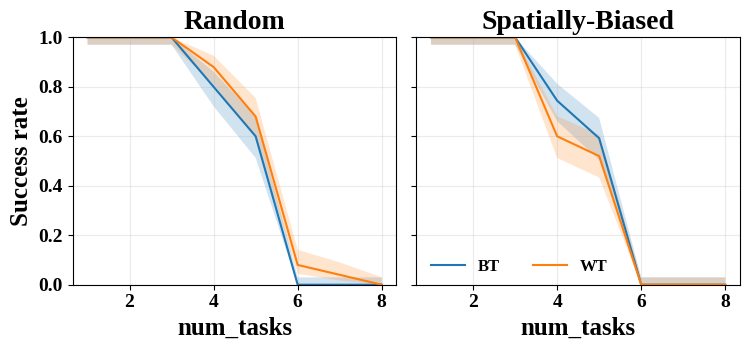}
  \caption{Success rates of suboptimal task selectors (BT vs. WT) on \textbf{random} (left) and \textbf{spatially-biased} (right) scenarios. Lower is better.}
  \label{fig:subopt_success_rate_selector_random/heatmap}
  \Description{Two side-by-side line graphs showing the success rates of the BT versus WT suboptimal task selectors. The left graph is for the `Random' scenario, and the right graph is for the `Spatially-Biased' scenario. In both graphs, the x-axis is the number of tasks and the y-axis is the success rate. The lines for BT and WT are plotted with confidence intervals, showing their comparative performance under different conditions.}
\end{figure}

The comparison of success rate is more nuanced. While pooling all scenarios shows no statistically significant overall difference between \texttt{WT} and \texttt{BT}, the stratified analysis reveals a consistent pattern shown in Figure.~\ref{fig:subopt_success_rate_selector_random/heatmap}: on the \emph{random} set, \texttt{WT} attains higher success rates across task counts, with Wilson 95\% CIs typically above \texttt{BT}; on the \emph{spatially-biased} set, \texttt{BT} prevails. In short, success favours \texttt{WT} on random layouts and \texttt{BT} on spatially clustered layouts, implying the scenario-sensitive property of different task selectors.

Finally, we synthesise our findings in a holistic comparison of the runtime-solution quality trade-off. We select \texttt{WT} as the representative task-centric solver for this comparison, as our prior analysis established its superior solution quality over \texttt{BT} while maintaining comparable success rates. We filtered out trivial instances with fewer than 3 tasks to ensure the analysis captures meaningful performance distinctions, resulting in 372 commonly solved scenarios across all five algorithm families. Table~\ref{tab:avg-ranks-runtime-gap} reveals a clear performance spectrum. The \texttt{Optimal} variants provide the highest quality solutions at the greatest computational cost, while the heuristic \texttt{Greedy-PP} is the fastest but yields the poorest solutions. Our proposed task-centric solvers (\texttt{WT}) and the agent-centric baselines (\texttt{-nn}) occupy the intermediate space, where a significant trade-off is confirmed (Friedman: $W=0.722,\ p \approx 10^{-231}$ for runtime; $W=0.703,\ p \approx 10^{-134}$ for solution gap). Specifically, pairwise post-hoc tests show that our solvers are significantly faster than the \texttt{-nn} baselines, which in turn produce higher-quality solutions. This analysis positions our task-centric solvers as a new, more efficient frontier on the runtime-quality spectrum.

\begin{table}[t]
  \centering
  \small
  \setlength{\tabcolsep}{3.5pt}
  \caption{Comparison of runtime and solution gap (mean $\pm$ std) along with their average ranks across 372 common instances. Lower is better. CT-CBS-WT is abbreviated as WT.}
  \label{tab:avg-ranks-runtime-gap}
  \label{tab:avg-ranks-runtime-gap}
  \begin{tabular}{lcccc}
    \toprule
    \multirow{2}{*}{\textbf{Algorithm}} & \multicolumn{2}{c}{\textbf{Runtime}} & \multicolumn{2}{c}{\textbf{Solution Gap}} \\
    \cmidrule(lr){2-3} \cmidrule(lr){4-5}
     & \textbf{Time (s)} & \textbf{AvgRank} & \textbf{Gap (\%)} & \textbf{AvgRank} \\
    \midrule
    \texttt{Optimal}   & $203.30 \pm 201.75$ & 4.36 & $\mathbf{0.00} \pm \mathbf{0.00}$ & \textbf{1.00} \\
    \texttt{NN2}       & $186.69 \pm 197.58$ & 4.01 & $0.05 \pm 0.49$ & 1.02 \\
    \texttt{NN1}       & $133.12 \pm 173.47$ & 2.95 & $1.32 \pm 2.90$ & 1.72 \\
    \texttt{WT}        & $107.33 \pm 165.36$ & 2.59 & $3.47 \pm 5.81$ & 2.18 \\
    \texttt{Greedy-PP} & $\mathbf{0.03} \pm \mathbf{0.01}$ & \textbf{1.00} & $24.06 \pm 22.20$ & 4.55 \\
    \bottomrule
  \end{tabular}
\end{table}


\section{Conclusion and Future Work}

In this paper, we formalised a novel Cooperative Transportation Task Allocation and Path Finding (CT-TAPF) problem and introduced CT-TCBS, a provably optimal solver. We demonstrated that our incremental expansion strategy is critical for performance and identified a `task-conflict expansion dilemma' where sophisticated pathfinding resolvers can be detrimental in this integrated setting. To address the computational cost of optimality, we also developed a family of suboptimal solvers (CT-TCBS-BT/WT) that adopt a global, task-centric perspective. Our experiments show these suboptimal methods establish a new, more efficient frontier on the solution quality-runtime spectrum. Future research will extend this framework to continuous MAPD within realistic warehouses, exploring decentralised control via social auctions and negotiation. Additionally, introducing heterogeneous agents with specialised capabilities could further enhance system flexibility and search efficiency.



\begin{acks}
This work was supported by the China Scholarship Council (CSC No. 202408060197). Generative AI tools were used for language editing and code generation; the authors verified all content and assume full responsibility for this publication.
\end{acks}



\bibliographystyle{ACM-Reference-Format} 
\bibliography{sample}


\end{document}